\documentclass[lettersize,journal]{IEEEtran}
\usepackage{amsmath,amsfonts}
\usepackage{algorithmic}
\usepackage{algorithm}
\usepackage[justification=raggedright,singlelinecheck=false]{caption}
\usepackage{amsthm}
\usepackage{amssymb}
\usepackage{array}
\usepackage{amsfonts}
\usepackage{afterpage}
\usepackage[caption=false,font=normalsize,labelfont=sf,textfont=sf]{subfig}
\usepackage{textcomp}
\usepackage{stfloats}
\usepackage{url}
\usepackage{verbatim}
\usepackage{graphicx}
\usepackage{cite}
\usepackage{ragged2e}
\usepackage{caption}
\usepackage{float}
\usepackage{xcolor}
\usepackage{amsmath}

\begin{document}

\title{OFDM-Based Active STAR-RIS-Aided Integrated Sensing and Communication Systems}

\author{Hanxiao Ge,~\IEEEmembership{Graduate~Student~Member,~IEEE,} Anastasios~Papazafeiropoulos,~\IEEEmembership{Senior~Member,~IEEE,}\\and~Tharmalingam~Ratnarajah,~\IEEEmembership{Senior~Member,~IEEE}
\thanks{H. Ge, and T. Ratnarajah are with Institute for Imaging, Data and Communications, The University of Edinburgh, Edinburgh, EH9 3BF, UK (e-mails: \{hanxiao.ge, t.ratnarajah\}@ed.ac.uk).}
\thanks{A. Papazafeiropoulos is with the Communications and Intelligent
Systems Research Group, University of Hertfordshire, Hatfield AL10 9AB,
U. K., and with SnT at the University of Luxembourg, Luxembourg (email: tapapazaf@gmail.com).}}

\maketitle

\begin{abstract}
 Simultaneously transmitting and reflecting reconfigurable intelligent surface (STAR-RIS), which consists of numerous passive elements, has recently emerged in wireless communication systems as a promising technology providing 360$^\circ$ coverage and better performance. In our research, we introduce an active STAR-RIS (ASTARS)-aided integrated sensing and communications (ISAC) system designed to optimize the radar signal-to-noise ratio (SNR), enhancing detection and signal transmission efficiency. The introduction  of an ISAC system aims to improve both communication
efficiency and sensing capabilities. Also, we employ orthogonal frequency division multiplexing (OFDM) to address the frequency-selective fading problem. Furthermore, we evaluate the radar sensing capabilities by examining the range and velocity, and assess the performance through the mean-squared error (MSE) of their estimations. Our simulation results demonstrate that ASTARS outperforms STAR-RIS in our system configurations, and that the proposed optimization approach further enhances the system performance. Additionally, we confirm that an increase in the subcarrier spacing can reduce the transmission bit error rate (BER) under high-velocity conditions.
\end{abstract}

\begin{IEEEkeywords}
Active RIS, STAR-RIS, ISAC, OFDM, 6G Communications
\end{IEEEkeywords}

\section{Introduction}
Reconfigurable intelligent surface (RIS), which deploy a large number of reflective elements, plays an important role in wireless communications. In \cite{0008} and \cite{0033}, authors have introduced that the RIS can reflect the beams to a suitable direction and improve the spectral efficiency (SE). Also, authors in \cite{0057} and \cite{0051} confirm that RIS can adjust its signals to improve connection quality, making communication more reliable and secure. In parallel, when there is a weak or no direct link for sensing, RIS can create a virtual link to fill in this gap \cite{0061,0062}.

However, conventional RIS is passive, and its effectiveness decreases because of the double fading effect \cite{0059,0060}. This effect represents the combined path loss when linking the transmitter to the receiver via the RIS, resulting from the combined losses of both the transmitter-RIS and RIS-receiver connections. In \cite{0076}, authors proposed an RIS architecture called active RIS to overcome the double fading effect of RIS-aided systems. The main characteristic of active RISs is their capability to reflect signals with active amplification. This can be achieved by incorporating reflection-type amplifiers into their reflecting components. By using more power, active RIS can offset the path loss of reflected links. This offers a potential solution to mitigate the double fading effect.

Unfortunately, in the case of  traditional passive and active RISs,  all transceivers and sensing targets are on the same side, which results in coverage limited to half the space. In \cite{0048,0053,0054}, authors have introduced the technology of simultaneously transmitting and reflecting RIS (STAR-RIS), which is of practical interest and improves the coverage and the SE compared to traditional RIS-aided systems. Authors in \cite{0044,0048} have examined the performance of rate-splitting in STAR-RIS-aided systems. Also, in \cite{0058}, the impact of phase noise in STAR-RIS-aided systems was studied.  Henceforth, active elements with STAR-RIS will be referred as active STAR-RIS (ASTARS). ASTARS
has been employed in previous studies, such as \cite{0066} and \cite{0080},
demonstrating its capability to achieve comprehensive spatial
coverage and effectively reduce the double fading effect.

Generally, RIS has been used in many network architectures, such as massive multiple-input multiple-output (mMIMO) systems, and integrated sensing and communication (ISAC) systems. The core concept of ISAC is to enable a node to carry out both radar sensing and wireless communication tasks using a single time-frequency-power-hardware resource.  In \cite{0070,0075,0077}, authors have considered the performance of RIS-aided ISAC networks. Additionally, when integrated into an ISAC system, RIS brings additional versatility. Not only does it enhance communication but also elevates the sensing capabilities of the system. This makes ISAC more adaptive and responsive to various environmental conditions. Also, the works \cite{0068,0071} have employed the active RIS in an ISAC network. Authors have confirmed that the radar signal to interference and noise ratio (SINR) of active RIS-aided ISAC systems perform better than the traditional RIS case. In the works, referenced as \cite{0073,0074,0093}, the authors utilized ISAC within STAR-RIS-aided systems, positioning the target and users on opposite sides of the STAR-RIS. The authors in \cite{0074} positioned both users and the target on the same side to satisfy the general conditions. However, this placement method may negatively impact and interfere with the sensing performance. 

In most RIS-aided ISAC research such as \cite{0070,0077}, authors primarily focus on maximizing the radar signal-to-noise ratio (SNR). However, they often overlook crucial radar sensing parameters such as the range and velocity. These parameters have been widely discussed in previous works, e.g., \cite{0069,0078}, which demonstrate the performance of sensing. In our work, we not only consider these parameters but also investigate the mean-squared error (MSE) of their estimation. Moreover, for most works about RIS-aided ISAC systems, authors only consider that the system has only a single subcarrier. However, the challenge of addressing frequency-selective fading in channels with a single subcarrier system is compounded by the highly intricate signal equalization required, making it impractical to maintain high data rates. Consequently, employing orthogonal frequency division multiplexing (OFDM) becomes essential to effectively tackle this issue. OFDM is commonly used in systems that combine radar and communications \cite{0049,0051,0005}. This method is advantageous because it makes a better use of the available frequency spectrum and is less affected by signal weakening over different frequencies, which is the key for effectively merging radar and communication systems. Only the authors in \cite{0075} have considered OFDM in the context of RIS-aided ISAC systems. However, their analysis is confined to the use of traditional passive RIS. There has been no research studying ASTARS-aided
ISAC systems.

In previous works, such as \cite{0079}, authors have only utilized low values of OFDM subcarrier spacing. However, this approach is not suitable for high-speed targets, as it leads to inaccurate data estimation. In the context of fifth-generation new radio (5G NR) and beyond, it is crucial to consider higher subcarrier spacing to mitigate the negative effects associated with high velocities. Reference \cite{0090} demonstrated that increased subcarrier spacing can reduce transmission errors. 

\subsection{Contributions}

Based on the aforementioned observations and gaps, in this work, we integrate the ASTARS with ISAC,
arranging users and the target on opposite sides of the RIS
panel. This system structure is designed to diminish the typical
interference that user communications exert on sensing. By combining the ASTARS in an ISAC framework, we propose a
novel approach to enhance system performance, ensuring more
reliable communication and accurate sensing by effectively
managing the interference between these two operations. Moreover, we integrate OFDM into the ASTARS-aided ISAC framework, to enhance the system capabilities. On this ground, we optimize the ASTARS phases and amplitudes along with transmit beamforming to maximize the radar SNR, thereby improving the overall performance. Through simulations, we assess the performance of the ASTARS and STAR-RIS assisted systems. Range and velocity estimations are carried out by using the optimized ASTARS configuration. Additionally, our work investigates the effects of varying OFDM subcarrier spacing, in alignment with 5G NR standards, with respect to the bit error rate (BER). Our main contributions are:

\begin{itemize}
    \item We propose an ASTARS-aided ISAC system, where the target and users are located on different sides. In the proposed system, the dual-function radar communication (DFRC) base station (BS) transmits a signal towards the ASTARS. Then, the ASTARS directs the beam towards the intended target. Upon receiving the beam, the target reflects the incoming signal back along its path, ultimately redirecting it towards the ASTARS. The ASTARS, once again, captures and reflects the signal back to the DFRC BS. Meanwhile, the BS transmits signals to users within the communication coverage area via the ASTARS. In our simulations, we have confirmed that ASTARS outperforms traditional STAR-RIS in enhancing the radar SNR.
    \item We derive equations for the radar SNR and communication SINR to optimize our ISAC system. By focusing on maximizing radar SNR, we significantly enhance radar sensing capabilities. Concurrently, we ensure that the communication aspect of our system meets the minimum SINR requirements for each user, through careful design of the precoder and the meticulous adjustment of ASTARS phases and amplitudes. This approach ensures that both sensing and communication work well together without sacrificing the quality of either function. 
Our simulations prove that this optimization works well, showing that our system performs better and is more reliable than traditional methods.
    \item We employ OFDM to tackle the effect of frequency-selective fading. By employing multiple subcarriers within the OFDM framework, we distribute a high data rate signal into multiple lower data rate signals that are concurrently transmitted across various frequencies. This approach significantly enhances the functionality and efficiency in radar sensing and wireless communications, surpassing the performance of previous RIS-aided ISAC systems that relied on a single subcarrier \cite{0077}.
    \item We employ the optimized ASTARS-aided system for range and velocity estimations. The simulation results indicate that as the true values of range and velocity increase, the MSE estimation also increases, leading to greater inaccuracies. Conversely, when the OFDM symbol encompasses more subcarriers, the MSE diminishes, resulting in more accurate estimations. This suggests that the accuracy of our system estimations is affected by the inherent challenges associated with larger ranges and higher velocities, as well as by the critical role of subcarrier density in enhancing the estimation precision.
    \item We consider the transmission BER in our simulations. Our findings indicate that as the actual velocity increases, the BER correspondingly rises. Furthermore, we have verified that an increase in subcarrier spacing leads to a reduction in BER. This evidence supports the conclusion that the system performance under high-velocity conditions can be enhanced by enlarging the subcarrier spacing.
\end{itemize}
\subsection{Paper outline}
For the rest of this paper, Section II introduces the structure of the ASTARS-aided ISAC system and the expressions of radar and communication signals. In Section III, we introduce the optimization problem of our system and the design of transmit beamforming, as well as the optimization of the ASTARS phases and amplitudes. We simplify the problem to enable the use of CVX for carrying out the optimization. 
In Section IV, we introduce the procedures for range and velocity estimations by using our optimized transmit beamforming and ASTARS phases and amplitudes. Section V includes simulations, and the Conclusion is presented in Section VI.

\subsection{Notations}
Scalars, vectors, and matrices are represented by lower case, lower case bold face, and upper case bold face letters, respectively. The notations $(.)^H$, $(.)^T$, and $\textup{tr}(.)$ denote the Hermitian, transpose, and trace operators, respectively. The conjugate is given as $(.)^{*}$. The $l_2$ norm is represented by $\|.\|$. The notation diag ($\mathbf{a}$)
denotes a matrix with elements equal to the diagonal elements
of $\mathbf{a}$. The expectation operator is denoted by $\mathbb{E}\{.\}$. Also, the vector notation is explicitly denoted by the $\textup{vec}$ symbol, as in $\textup{vec}(\mathbf{a})$ for vectors. The notation for real and imaginary values are $\mathbb{R}\{.\}$ and $\mathbb{I}\{.\}$. The symbol $\arg$ denotes the argument function.

\section{System models}
\subsection{ASTARS model}

 We consider a downlink ASTARS-aided ISAC system designed for simultaneous radar sensing and communication as Fig. $1$ \footnote{\textcolor{black}{Placing the target and users on opposite sides of the STAR-RIS allows simultaneous optimization of radar sensing and communication performance, leveraging the reflective and transmissive properties of the surface for enhanced system efficiency, see \cite{0073,0093}.}}. The DFRC BS transmits a signal to the ASTARS with $Q$ elements, which then directs the beam towards a target in the sensing area. Upon receiving the beam, the target reflects it back along its path, which is then captured by the ASTARS and redirected to the DFRC BS for processing. \textcolor{black}{The DFRC BS employs uniform linear arrays (ULAs) for both sensing and communication with $M$ antennas.} These arrays are utilized to communicate with $K$ users, each equipped with a single antenna. Notably, these users are situated on the opposite side of the ASTARS, distinct from the sensing target location. We assume that $\mathcal{K}_{t}=\{1,\ldots,K \} $ users are located on
	the communication region $ (t) $, i.e., behind the surface. We assume that all surface elements serve simultaneously all users in $t$ region and the target in $r$ region. The amplitude and phase parameters for the ASTARS elements are defined as follows: The amplitude for $q$th element $\beta_q^{i}$ lies within the range $[0,1]$ and the phase $\phi_q^{i}$ within $[0,2\pi)$, where the index $i$ denotes either communication ($t$) or sensing ($r$). The entire phase matrix $\boldsymbol{\Psi}_{i}$ is constructed as a diagonal matrix with elements $\alpha_q\beta_{q}^{i}e^{j\phi_{q}^{i}}$, resulting in $\boldsymbol{\Psi}_{i} = \mathrm{diag}(\alpha_1\beta_{1}^{i}e^{j\phi_{1}^{i}}, \ldots, \alpha_Q\beta_{Q}^{i}e^{j\phi_{Q}^{i}}) \in \mathbb{C}^{Q\times Q}$. The vector $\boldsymbol{\phi}_i$ contains the phase elements of each ASTARS element, expressed as $\boldsymbol{\phi}_i = [\alpha_1\beta_{1}^{i}e^{j\phi_{1}^{i}}, \ldots, \alpha_Q\beta_{Q}^{i}e^{j\phi_{Q}^{i}}]^T$. It is subject to the conditions that for each element $q$, the sum of the squares of the amplitudes for communication and sensing must be equal to one, i.e., ($(\beta_q^t)^2+(\beta_q^r)^2=1$), the active amplifying coefficient (AAC) $\alpha_q$ should not exceed a maximum value $\alpha_{max}$, and the phase $\phi_{q}^{i}$ must fall within the range $[0,2\pi)$ for all $q$. 
To acquire the OFDM symbol, the transmitted block signals undergo an inverse discrete Fourier transform (IDFT). Following the IDFT, the output is cyclically shifted to mitigate intersymbol interference (ISI) and to incorporate a guard interval. Once the signals traverse multipath channels, it becomes essential to eliminate the guard interval and execute the discrete Fourier transform (DFT) \cite{0005}, \cite{0085}. The assumption here is that the cyclic prefix (CP) length is sufficiently large to disregard any timing and frequency discrepancies.

\subsection{Transmitted signal from the DFRC BS}

The DFRC BS utilizes the transmit beamforming matrices $\mathbf{W}_r$ and $\mathbf{W}_c$ for radar and communication purposes, respectively. This configuration allows the DFRC BS to simultaneously undertake radar sensing and communication tasks. The combined transmitted signal for the $n$th subcarrier is given by 
 \begin{equation}
 \begin{aligned}
      \mathbf{x}(n)&=\mathbf{W}_r\mathbf{s}_r(n)+\mathbf{W}_c\mathbf{s}_c(n)\\ 
      &=\sum_{m=1}^M\mathbf{w}_{r,m}s_{r,m}(n)+\sum_{k=1}^K\mathbf{w}_{c,k}s_{c,k}(n)\\  &=\mathbf{W}\mathbf{S}(n),\label{eq1}
 \end{aligned}
 \end{equation}

\noindent where $\mathbf{s}_r(n)=[s_{r,1}(n),\ldots,s_{r,M}(n)]\in\mathbb{C}^{M\times 1}$ denotes the radar signal and $\mathbf{s}_c(n)=[s_{c,1}(n),\ldots,s_{c,K}(n)]\in\mathbb{C}^{K\times 1}$ denotes the transmitted symbol to $K$ users. The transmit beamforming matrices  $\mathbf{W}_r$ and $\mathbf{W}_c$ can be written as $\mathbf{W}_r=[\mathbf{w}_{r,1},\mathbf{w}_{r,2},\ldots,\mathbf{w}_{r,M}]\in\mathbb{C}^{M\times M}$ and $\mathbf{W}_c=[\mathbf{w}_{c,1},\mathbf{w}_{c,2},\ldots,\mathbf{w}_{c,K}]\in\mathbb{C}^{M\times K}$. Also, we have $\mathbf{W}=[\mathbf{W}_r,\mathbf{W}_c]\in\mathbb{C}^{M\times (M+K)}$. The radar and communication signals are uncorrelated and the transmitted signal covariance can be written as
\begin{equation}
    \mathbf{R}=\mathbb{E}\{\mathbf{x}\mathbf{x}^H\}=\mathbf{W}_r\mathbf{W}_r^H+\sum_{k=1}^K \mathbf{R}_k,
\end{equation}

\noindent where we have $\mathbf{R}_k=\mathbf{w}_{c,k}\mathbf{w}_{c,k}^H\in\mathbb{C}^{M\times M}$. This expression shows the independence of radar and communication functionalities, with $\mathbf{R}$ encapsulating the aggregate effect on signal covariance, enabling the efficient utilization of the DFRC system dual capabilities.

\begin{figure}[t]
\flushleft
\includegraphics[width=0.490\textwidth]{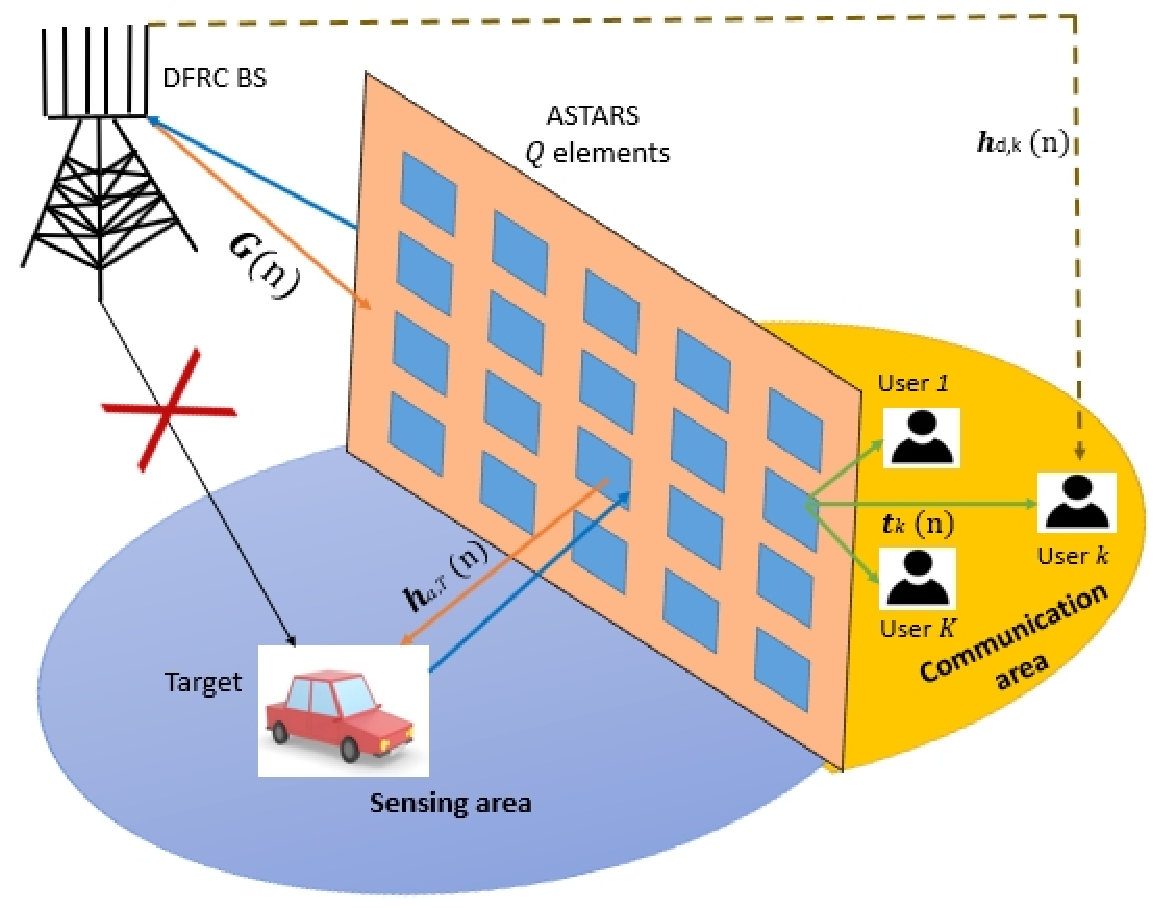}
\caption{Structure of the ASTARS-aided ISAC system.}
\end{figure}
\subsection{Signal at the radar receiver}

For the channels in the radar sensing area, we consider that the channel between DFRC BS and the ASTARS is $\mathbf{G}(n)\in\mathbb{C}^{Q\times M}$, which is based on Rician fading with the LoS part and the Rician factor is $R$. The expression of $\mathbf{G}(n)$ can be written as 
\textcolor{black}{\begin{equation}
    \mathbf{G}(n)=\sqrt{\frac{R}{R+1}}\mathbf{G}_{\textup{LoS}}(n)+\sqrt{\frac{1}{R+1}}\mathbf{G}_{\textup{nLoS}}(n),
\end{equation}}
\noindent where $\mathbf{G}_{\textup{LoS}}$ and $\mathbf{G}_{\textup{nLoS}}$ are LoS and nLoS parts of the channel $\mathbf{G}$. The LoS $\mathbf{G}_{\textup{LoS}}(n)$ can be written by using steering vectors as $\mathbf{G}_{\textup{LoS}}(n)=\mathbf{a}_q(\theta_0)\mathbf{a}_m^H(\theta_0)\in\mathbb{C}^{Q\times M}$ with $\mathbf{a}_q(\theta_0) =[1,e^{-j\pi\textup{sin}(\theta_0)},\ldots,e^{-j\pi(Q-1)\textup{sin}(\theta_0)}]^T$ and $\mathbf{a}_m(\theta_0) =[1,e^{-j\pi\textup{sin}(\theta_0)},\ldots,e^{-j\pi(M-1)\textup{sin}(\theta_0)}]^T$. The angle $\theta_0$ is from $-\frac{\pi}{2}$ to $\frac{\pi}{2}$. The channel between the ASTARS and the target is $\mathbf{h}_{a,T}(n)\in\mathbb{C}^{Q\times 1}$. To simpify the complexity, the direct link from the DFRC BS to the target is blocked \footnote{\textcolor{black}{Isolating the DFRC BS from the direct link to the target enhances system security and reduces interference, allowing for more accurate radar sensing and communication through the ASTARS, see \cite{0071}.}}. Hence, the collected echo signal at the DFRC BS from the target can be written as
\begin{equation}
\begin{aligned}
    &\mathbf{y}_r(n)=\mathbf{G}^H(n)[\boldsymbol{\Psi}_r\mathbf{h}_{a,T}(n)\mathbf{h}_{a,T}^H(n)\boldsymbol{\Psi}_r\mathbf{G}(n)(\mathbf{x}(n)+\mathbf{v}_1(n))\\
    &+\mathbf{v}_2(n)]+\mathbf{n}_r(n)\\
    &=\mathbf{G}^H(n)\boldsymbol{\Psi}_r\mathbf{h}_{a,T}(n)\mathbf{h}_{a,T}^H(n)\boldsymbol{\Psi}_r\mathbf{G}(n)\mathbf{x}(n)\\
    &+\mathbf{G}^H(n)\boldsymbol{\Psi}_r\mathbf{h}_{a,T}(n)\mathbf{h}_{a,T}^H(n)\boldsymbol{\Psi}_r\mathbf{v}_1(n)+\mathbf{G}^H(n)\boldsymbol{\Psi}_r\mathbf{v}_2(n)\\
    &+\mathbf{n}_r(n)\\
    &=\mathbf{H}_T(n)\mathbf{x}(n)+\mathbf{H}_{v,1}(n)\mathbf{v}_1(n)+\mathbf{H}_{v,2}(n)\mathbf{v}_2(n)+\mathbf{n}_r(n),\label{eq3}
\end{aligned} 
\end{equation}

\noindent where $\mathbf{n}_r(n)\in \mathcal{CN}(\mathbf{0},\sigma_r^2\mathbf{I}_M)$  is the additive white Gaussian noise (AWGN) at the DFRC BS. The noise vector $\mathbf{v}_1(n)\in \mathcal{CN}(\mathbf{0},\sigma_v^2\mathbf{I}_Q)$ is the dynamic noise at the ASTARS. Also, the dynamic noise introduced by the active RIS upon the return of the radar signal is represented by $\mathbf{v}_2(n)$, which is independent and identically distributed as $\mathbf{v}_1(n)$. In the above equation, we have $\mathbf{H}_T(n)=\mathbf{G}^H(n)\boldsymbol{\Psi}_r\mathbf{h}_{a,T}(n)\mathbf{h}_{a,T}^H(n)\boldsymbol{\Psi}_r\mathbf{G}(n)$, $\mathbf{H}_{v,1}(n)=\mathbf{G}^H(n)\boldsymbol{\Psi}_r\mathbf{h}_{a,T}(n)\mathbf{h}_{a,T}^H(n)\boldsymbol{\Psi}_r$ and $\mathbf{H}_{v,2}(n)=\mathbf{G}^H(n)\boldsymbol{\Psi}_r$. When considering the ASTARS-to-target channels for each OFDM subcarrier, \textcolor{black}{we represent them as $\mathbf{h}_{a,T}(n)$. These channels are characterized by employing the concept of a steering vector, denoted by $\mathbf{a}(\theta_{aT})$ as
\begin{align}
&\mathbf{h}_{a,T}(n)=\sqrt{\widetilde{\beta}_a}\mathbf{a}(\theta_{aT}),
\end{align}}

\noindent where $\widetilde{\beta}_a$ is the large-scale fading coefficient of the channel $\mathbf{h}_{a,T}(n)$. Also, the steering vector can be written as 
\begin{equation}
    \mathbf{a}(\theta_{aT}) =[1,e^{-j\pi\textup{sin}(\theta_{aT})},\ldots,e^{-j\pi(Q-1)\textup{sin}(\theta_{aT})}]^T,
\end{equation}

\noindent where $\theta_{aT}\in [-\frac{\pi}{2},\frac{\pi}{2}]$. From \eqref{eq1} and \eqref{eq3}, we can write the radar SNR by calculating the power of each part in Eq. \eqref{eq1} as \eqref{eq6}.
\begin{figure*}[t]
\begin{equation}
    \textup{SNR}_r(n) = \frac{\mathrm{tr}(\mathbf{H}_T(n)\mathbf{W}\mathbf{W}^H\mathbf{H}_T^H(n))}{\mathrm{tr}(\sigma_v^2\mathbf{H}_{v,1}(n)\mathbf{H}_{v,1}^H(n)+\sigma_v^2\mathbf{H}_{v,2}(n)\mathbf{H}_{v,2}^H(n)+\sigma_r^2\mathbf{I}_M)}\label{eq6}
\end{equation}
  \hrulefill
\end{figure*}
\subsection{Communication signal at the users}
For the communication process, the signal from the DFRC BS is sent to users either directly or via an indirect route involving ASTARS, classifying these as direct or indirect links respectively. Unlike radar sensing, we maintain a direct link within the communication area. In the direct link $\mathbf{h}_{d,k}(n)$, the signal is directly transmitted from the DFRC BS to the users without any reflections. Conversely, in the indirect link, the signal is initially relayed from the DFRC BS to ASTARS and subsequently forwarded to the users. The collective received signal at the communication user $k$ can be expressed as
\begin{equation}
\begin{aligned}
        &y_{c,k}(n)\\
        &=(\mathbf{t}_k^H(n)\boldsymbol{\Psi}_{t}^H(n)\mathbf{G}(n)+\mathbf{h}_{d,k}^H(n))\mathbf{x}(n)+\mathbf{t}_k^H(n)\boldsymbol{\Psi}_{t}^H\mathbf{v}_1(n)\\
        &+n_{k}(n)\\
        &=\underbrace{\mathbf{u}_k^H(n)\mathbf{w}_{c,k}s_{c,k}(n)}_{\textup{desired signal}}+\underbrace{\mathbf{u}_k^H(n)\mathbf{W}_r\mathbf{s}_r(n)}_{\textup{sensing interference}}+\underbrace{\sum_{i\neq k}\mathbf{u}_k^H(n)\mathbf{w}_{c,i}s_{c,i}}_{\textup{cross-interference}}\\
        &+\underbrace{\mathbf{t}_k^H(n)\boldsymbol{\Psi}_{t}^H\mathbf{v}_1(n)}_{\textup{dynamic noise}}+n_{k}(n),
\end{aligned}
\end{equation}

\begin{figure*}[t]

    \begin{equation}
        \mathrm{SINR}_{k}(n) = \frac{|\mathbf{u}_k^H(n)\mathbf{w}_{c,k}|_2^2}{|\mathbf{u}_k^H(n)\mathbf{W}_r|_2^2+\sum_{i\neq k}\mathbf{u}_k^H(n)\mathbf{R}_i\mathbf{u}_k(n)+\sigma_v^2\mathbf{t}_k^H(n)\boldsymbol{\Psi}_{t}\boldsymbol{\Psi}_{t}^H(n)\mathbf{t}_k(n)+\sigma^2}
        \label{eq8}
    \end{equation}
     \hrulefill
\end{figure*}

\noindent where $\mathbf{h}_{d,k}(n)\in\mathbb{C}^{M\times 1}$ is the direct link from DFRC BS to users in the $n$th subcarrier; $\mathbf{t}_k(n)\in\mathbb{C}^{Q\times 1}$ is the channel between the STAR-RIS and user $k$.  $n_{k}(n)\sim \mathcal{CN}(0,\sigma^2)$ is the AWGN. The SINR for the $k$th user can be given as \eqref{eq8}, where $\mathbf{u}_k(n)=\mathbf{h}_{d,k}(n)+\mathbf{G}^H(n)\boldsymbol{\Psi}_{t}\mathbf{t}_k(n)$ is the overall channel between users and the DFRC BS. Also, channels $\mathbf{h}_{d,k}$ and $\mathbf{t}_k$ are based on Rayleigh fading, with large-scale fading coefficients $\widetilde{\beta}_{d,k}$, $\widetilde{\beta}_G$, and $\eta_k$.



\section{Problem formulation and proposed solution}

\subsection{Problem formulation}
From \eqref{eq6} and \eqref{eq8}, we can find that the transmit beamforming at the DFRC BS $\mathbf{W}$ and the transmission and reflection beamforming at the ASTARS ($\boldsymbol{\Psi}_t$ and $\boldsymbol{\Psi}_r$) are shared between the dual functions. We need to design them jointly to maximize the radar SNR and to meet the minimum SINR for each communication users \cite{0093}. This balance is crucial because maximizing the radar SNR enhances target detection capabilities, and maintaining a minimum SINR for communication ensures a reliable data transmission. The problem of finding the optimal beamforming configurations to meet these objectives can be formally stated as a joint optimization problem as

\begin{subequations}
\begin{align}
    \mathcal{P}1:&\max_{\mathbf{W},\boldsymbol{\Phi}_{i}} \quad\quad\quad\mathrm{SNR}_R(n)\nonumber\\
    &\quad s.t.\quad\,\,\quad\quad\textcolor{black}{\mathrm{SINR}_{k}(n)\geq \xi,\forall k}\label{9a}\\
    &\quad\quad\quad\quad\quad\quad\textcolor{black}{\mathrm{tr}(\mathbf{W}\mathbf{W}^H)\leq P_{BS}}\label{9b}\\
    &\quad\quad\quad\quad\quad\quad(\beta_q^t)^2+(\beta_q^r)^2=1,q=1,\ldots,Q\label{9c}\\
    &\quad\quad\quad\quad\quad\quad\beta_q^t, \beta_q^r\in [0,1], q=1,\ldots,Q\label{9d}\\
 &\quad\quad\quad\quad\quad\quad\phi_q^t, \phi_q^r\in [0,2\pi), q=1,\ldots,Q\label{9e}\\
 &\quad\quad\quad\quad\quad\quad \alpha_q\leq \alpha_{\textup{max}},\label{9f}
\end{align}
\end{subequations}

\noindent where $\xi$ is the SINR threshold for the $k$th user. Also, $P_{\mathrm{BS}}$ is the maximum trasmit power for the DFRC BS. It is clear that $\mathcal{P}1$ is non-convex and hard to be solved because of the complicated $\mathrm{SNR}_R$ equation. To solve this problem, we separate $\mathcal{P}1$ into several tractable sub-problems. One sub-problem involves with the optimization of the transmit beamforming, while another one focuses on the optimization of the ASTARS phase coefficients.

\subsection{Transmit beamforming $\mathbf{W}$ optimization}
To solve $\mathcal{P}1$, we can write the sub-problem $\mathcal{P}2$ to optimize the transmit beamforming $\mathbf{W}$
firstly. We keep $\boldsymbol{\Psi}_r$ and $\boldsymbol{\Psi}_t$ fixed and then maximize the sensing SNR. 
Then, the problem $\mathcal{P}2$ can be written as

\begin{equation}
\begin{aligned}
    \mathcal{P}2:&\max_{\mathbf{W}} \quad f(\mathbf{W})\\
    & s.t.\quad(1+\xi^{-1})\textup{tr}(\mathbf{u}_k(n)\mathbf{u}_k^H(n)\mathbf{R}_k)\geq\sigma^2\\
    &\quad+\textup{tr}(\mathbf{u}_k(n)\mathbf{u}_k^H(n)\mathbf{R}+\sigma_v^2\mathbf{t}_k^H(n)\boldsymbol{\Psi}_{t}\boldsymbol{\Psi}_{t}^H\mathbf{t}_k(n)),\forall k,\\
    &\;\quad\textcolor{black}{\mathrm{tr}(\mathbf{W}\mathbf{W}^H)\leq P_{BS}},\\
    &\;\quad \alpha_q \leq \alpha_{max},\forall q,\label{equ11}
\end{aligned}
\end{equation}

\noindent where the problem $\mathcal{P}2$ is a convex quadratic semidefinite program (QSDP) due to its quadratic objective and constraints that include linear equalities and inequalities with semidefinite conditions on the variables. The procedure for obtaining the constraint for the $\textup{SINR}_k$ has been given in Appendix $A$. The quadratic nature of the objective and the semidefinite constraints stemming from the beamforming matrices define it as a QSDP \cite{0086}. Such problems are solvable via convex optimization tools like CVX \cite{0088}, which use interior-point methods to achieve global solutions. This makes the approach good for handling the matrix variables and semidefinite constraints in beamforming optimization.

\subsection{\textcolor{black}{Phase coefficients optimization}}

When the transmit beamforming is obtained, we can formulate the problem $\mathcal{P}3$ to optimize the phase coefficients of ASTARS. We utilize two functions $g(\boldsymbol{\phi}_r,n)$ and $f(\boldsymbol{\phi}_r,n)$ to denote the complex terms in the radar SNR equation \cite{0071}. Consequently, the maximization problem can be reformulated as follows
\vspace{-0.25cm}
\begin{equation}
\begin{aligned}
    \mathcal{P}3:&\max_{\boldsymbol{\phi}_r,\boldsymbol{\phi}_t} \quad \textup{SINR}_R= \frac{g(\boldsymbol{\phi}_r,n)}{f(\boldsymbol{\phi}_r,n)}\\
    & s.t. \quad\quad\eqref{9a}, \eqref{9c}-\eqref{9f},
\end{aligned}
\end{equation}

\noindent where $g(\boldsymbol{\phi}_r,n)=\textup{tr}(\mathbf{H}_T(n)\mathbf{W}\mathbf{W}^H\mathbf{H}_T^H(n))$ and $f(\boldsymbol{\phi}_r,n)=\textup{tr}(\sigma_v^2\mathbf{H}_{v,1}(n)\mathbf{H}_{v,1}^H(n)+\sigma_v^2\mathbf{H}_{v,2}(n)\mathbf{H}_{v,2}^H(n)+\sigma_r^2\mathbf{I}_M)$. Problem $\mathcal{P}3$ is challenging because it is a fractional problem. We can use a new auxiliary valuable to rewrite the problem as a linear form and make it simple as
\vspace{-0.1cm}
\begin{equation}
    \max_{\boldsymbol{\phi}_r,\boldsymbol{\phi}_t} g(\boldsymbol{\phi}_r, n)-\delta f(\boldsymbol{\phi}_r, n),\label{eq14}
\end{equation}

\noindent where $\delta$ can be written in a close-form as $\delta=\frac{g(\boldsymbol{\phi}_r)}{f(\boldsymbol{\phi}_r)}$. This auxiliary variable aids in decomposing the original problem into a more solvable form, especially for the related optimizing parameters. For $\mathbf{H}_T(n)$, we can use the property $\boldsymbol{\Psi}_r\mathbf{h}_{a,T}(n)=\mathrm{diag}(\mathbf{h}_{a,T}(n))\boldsymbol{\phi}$ and then we can write $\overline{\mathbf{G}}(n)=\mathbf{G}^H(n)\mathrm{diag}(\mathbf{h}_{a,T}(n))$. Hence, $\mathbf{H}_T(n)$ can be rewritten as $\mathbf{H}_T(n)=\overline{\mathbf{G}}(n)\boldsymbol{\phi}_r\boldsymbol{\phi}_r^H\overline{\mathbf{G}}^H(n)$. According to these properties, $g(\boldsymbol{\phi}_r,n)$ can be further derived as
\begin{subequations}
    \begin{align}
      &g(\boldsymbol{\phi}_r,n)=\mathrm{tr}(\mathbf{H}_T(n)\mathbf{W}\mathbf{W}^H\mathbf{H}_T^H(n))\\
    &=  \mathrm{tr}(\overline{\mathbf{G}}(n)\boldsymbol{\phi}_r\boldsymbol{\phi}_r^H\overline{\mathbf{G}}^H(n)\mathbf{W}\mathbf{W}^H\overline{\mathbf{G}}(n)\boldsymbol{\phi}_r\boldsymbol{\phi}_r^H\overline{\mathbf{G}}^H(n))\\
    &=\mathrm{tr}(\overline{\mathbf{G}}^H(n)\overline{\mathbf{G}}(n)\boldsymbol{\phi}_r\boldsymbol{\phi}_r^H\overline{\mathbf{G}}^H(n)\mathbf{W}\mathbf{W}^H\overline{\mathbf{G}}(n)\boldsymbol{\phi}_r\boldsymbol{\phi}_r^H)\label{eq14c}\\
    &=\mathrm{vec}^H(\boldsymbol{\phi}_r\boldsymbol{\phi}_r^H)\big[(\overline{\mathbf{G}}^T(n)\mathbf{W}^*\mathbf{W}^T\overline{\mathbf{G}}^*(n))\nonumber\\
    &\otimes(\overline{\mathbf{G}}^H(n)\overline{\mathbf{G}}(n))\big]\mathrm{vec}(\boldsymbol{\phi}_r\boldsymbol{\phi}_r^H)\label{eq14d}\\
&=\mathbf{z}^H\mathbf{E}(n)\mathbf{z},\label{eq14e}
    \end{align}
\end{subequations}

\noindent where we have $\mathbf{z}=\mathrm{vec}(\boldsymbol{\phi}_r\boldsymbol{\phi}_r^H)$. From \eqref{eq14c} to \eqref{eq14d}, we have used $\mathrm{tr}(\mathbf{B}\mathbf{K}\mathbf{C}\mathbf{K})=\mathrm{vec}^H(\mathbf{K})\big[(\mathbf{C}^T\otimes\mathbf{B})\big]\mathrm{vec}(\mathbf{K})$. From \eqref{eq14d} to \eqref{eq14e}, we assume that $\mathbf{E}(n)=(\overline{\mathbf{G}}^T(n)\mathbf{W}^*\mathbf{W}^T\overline{\mathbf{G}}(n)^*)\otimes(\overline{\mathbf{G}}(n)^H\overline{\mathbf{G}}(n))$. Similarly, $f(\boldsymbol{\phi}_r, n)$ can be expressed as
\vspace{-0.1cm}
\begin{equation}
    \begin{aligned}
        &f(\boldsymbol{\phi}_r,n)\\
        &=\mathrm{tr}(\sigma_v^2\mathbf{H}_{v,1}(n)\mathbf{H}_{v,1}^H(n)+\sigma_v^2\mathbf{H}_{v,2}(n)\mathbf{H}_{v,2}^H(n)+\sigma_r^2\mathbf{I}_M),\label{equ15}
    \end{aligned}
\end{equation}

\noindent where we can write the above equation in detail. For the first term, we have 

\begin{equation}
    \begin{aligned}
        &\mathrm{tr}(\mathbf{H}_{v,1}(n)\mathbf{H}_{v,1}^H(n))\\
        &\overset{(a)}{=}\mathrm{tr}(\mathbf{G}^H(n)\boldsymbol{\Psi}_r\mathbf{h}_{a,T}(n)\mathbf{h}_{a,T}^H(n)\boldsymbol{\Psi}_r)\\
    &\overset{(b)}{=}\mathrm{tr}(\overline{\mathbf{G}}(n)\boldsymbol{\phi}_r\boldsymbol{\phi}_r^H\mathrm{diag}(\mathbf{h}_{a,T}^H(n))\mathrm{diag}(\mathbf{h}_{a,T}(n))\boldsymbol{\phi}_r\boldsymbol{\phi}_r^H\overline{\mathbf{G}}^H(n))\\
    &\overset{(c)}{=}\mathrm{tr}(\overline{\mathbf{G}}^H(n)\overline{\mathbf{G}}(n)\boldsymbol{\phi}_r\boldsymbol{\phi}_r^H\mathrm{diag}(\mathbf{h}_{a,T}^H(n))\mathrm{diag}(\mathbf{h}_{a,T}(n))\boldsymbol{\phi}_r\boldsymbol{\phi}_r^H)\\
    &\overset{(d)}{=}\mathrm{vec}^H(\boldsymbol{\phi}_r\boldsymbol{\phi}_r^H)\big[(\mathrm{diag}(\mathbf{h}_{a,T}^T(n))\mathrm{diag}(\mathbf{h}_{a,T}^*)(n))\\
    &\otimes(\overline{\mathbf{G}}^H(n)\overline{\mathbf{G}}(n))\big]\mathrm{vec}(\boldsymbol{\phi}_r\boldsymbol{\phi}_r^H)\\
    &\overset{(e)}{=}\mathbf{z}^H\mathbf{F}(n)\mathbf{z},\label{equ16}
    \end{aligned}
\end{equation}

\noindent where we obtain $(a)$ from \eqref{eq3}. In $(b)$, we use $\boldsymbol{\Psi}_r\mathbf{h}_{a,T}(n)=\mathrm{diag}(\mathbf{h}_{a,T}(n))\boldsymbol{\phi}$. Besides, from $(c)$ to $(d)$, we also use the property $\mathrm{tr}(\mathbf{B}\mathbf{K}\mathbf{C}\mathbf{K})=\mathrm{vec}^H(\mathbf{K})\big[(\mathbf{C}^T\otimes\mathbf{B})\big]\mathrm{vec}(\mathbf{K})$. In \eqref{equ16}, we have $\mathbf{F}(n)=(\mathrm{diag}(\mathbf{h}_{a,T}^T(n))\mathrm{diag}(\mathbf{h}_{a,T}^*(n)))\otimes(\overline{\mathbf{G}}^H(n)\overline{\mathbf{G}}(n))\big]\mathrm{vec}(\boldsymbol{\phi}_r\boldsymbol{\phi}_r^H)$. For the second term of \eqref{equ15}, it can be rewritten as
\begin{equation}
    \begin{aligned}
        &\mathbf{H}_{v,2}(n)\mathbf{H}_{v,2}^H(n)=\mathbf{G}^H(n)\boldsymbol{\Psi}_r\boldsymbol{\Psi}_r^H\mathbf{G}(n)\\
        &=\boldsymbol{\phi}_r^H\mathbf{G}(n)\mathbf{G}^H(n)\boldsymbol{\phi}_r=\boldsymbol{\phi}_r^H\mathbf{L}(n)\boldsymbol{\phi}_r,\label{equ17}
    \end{aligned}
\end{equation}

\noindent where in \eqref{equ17}, we have $\mathbf{L}(n)=\mathbf{G}(n)\mathbf{G}^H(n)$. Hence, the function $f(\boldsymbol{\phi}_r,n)$ can be rewritten by using \eqref{equ16} and \eqref{equ17} as

\begin{equation}
    \begin{aligned}
        &f(\boldsymbol{\phi}_r,n)\\
        &=\mathrm{tr}(\sigma_v^2\mathbf{H}_{v,1}(n)\mathbf{H}_{v,1}^H(n)+\sigma_v^2\mathbf{H}_{v,2}(n)\mathbf{H}_{v,2}^H(n)+\sigma_r^2\mathbf{I}_M)\\
        &=\sigma_v^2\mathbf{z}^H\mathbf{F}(n)\mathbf{z}+\sigma_v^2\boldsymbol{\phi}_r^H\mathbf{L}(n)\boldsymbol{\phi}_r+\sigma_r^2M.\label{equ18}
    \end{aligned}
\end{equation}

By inserting the equation \eqref{equ18} into \eqref{eq14}, the problem can be rewritten as

\begin{equation}
    \begin{aligned}
        \min_{\boldsymbol{\phi}_r,\boldsymbol{\phi}_t} &\delta f(\boldsymbol{\phi}_r,n)-g(\boldsymbol{\phi}_r,n)\\
        &=\delta(\sigma_v^2\mathbf{z}^H\mathbf{F}(n)\mathbf{z}+\sigma_v^2\boldsymbol{\phi}_r^H\mathbf{L}(n)\boldsymbol{\phi}_r+\sigma_r^2M)-\mathbf{z}^H\mathbf{E}(n)\mathbf{z}\\
        &=\mathbf{z}^H(\delta\sigma_v^2\mathbf{F}(n)-\mathbf{E}(n))\mathbf{z}+\sigma_v^2\delta\boldsymbol{\phi}_r^H\mathbf{L}(n)\boldsymbol{\phi}_r+\delta\sigma_r^2M\\
        &=\mathbf{z}^H\mathbf{M}(n)\mathbf{z}+\delta\sigma_v^2\boldsymbol{\phi}_r^H\mathbf{L}(n)\boldsymbol{\phi}_r+\delta\sigma_r^2M,\label{eq20}
    \end{aligned}
\end{equation}
\noindent where we change the problem to a minimization problem because it is clearly that $\delta f(\boldsymbol{\phi}_r,n)>g(\boldsymbol{\phi}_r,n)$. To further simpify \eqref{eq20}, we expand $\mathbf{z}^H\mathbf{M}(n)\mathbf{z}$ according to second-order Taylor expansion as \cite{0072}
\begin{equation}
    \begin{aligned}
        &\mathbf{z}^H\mathbf{M}(n)\mathbf{z}\\
        &\leq \lambda_{m,0}(n)\mathbf{z}^H\mathbf{z}+2\mathbb{R}(\mathbf{z}^H(\mathbf{M}(n)-\lambda_{m,0}(n)\mathbf{I}_{N^2})\mathbf{z}_j)\\
        &+\mathbf{z}_j^H(\lambda_{m,0}(n)\mathbf{I}_{N^2}-\mathbf{M}(n))\mathbf{z}_j,
    \end{aligned}
\end{equation}

\noindent where $\lambda_{m,0}(n)$ is the eigenvalue of $\mathbf{M}(n)$. $j$ means $j$th iteration for Majorization-Minimization (MM) algorithms \cite{0072,0091}. For the second term, we have 
\begin{equation}
\begin{aligned}
     &  2\mathbb{R}(\mathbf{z}^H(\mathbf{M}(n)-\lambda_{m,0}(n)\mathbf{I}_{N^2})\mathbf{z}_j)\\
        &=\mathbb{R}(\boldsymbol{\phi}^H\Gamma[(2\mathbf{M}(n)-2\lambda_{m,0}(n)\mathbf{I}_{N^2})\mathbf{z}_j]\boldsymbol{\phi})\\
        &=\mathbb{R}(\boldsymbol{\phi}^H\overline{\mathbf{M}}(n)\boldsymbol{\phi}),
\end{aligned}
\end{equation}

\noindent where $\Gamma$ is the reverse process of flattening a matrix into a vector. The above equation is still non-convex with respect to $\boldsymbol{\phi}$ due to its nature as a complex-valued function. It is necessary to transform $\mathbb{R}(\boldsymbol{\phi}^H\overline{\mathbf{M}}(n)\boldsymbol{\phi})$ into a corresponding real-valued function by using a symmetrical matrix $\widetilde{\mathbf{M}}$,  which can be expressed as \cite{0093}\\
\begin{align*}
&\widetilde{\boldsymbol{\phi}}_r^H\widetilde{\mathbf{M}}(n)\widetilde{\boldsymbol{\phi}}_r \\
&= 
\begin{bmatrix}
\mathbb{R}(\boldsymbol{\phi}_r^H), & \mathbb{I}(\boldsymbol{\phi}_r^H)
\end{bmatrix}
\begin{bmatrix}
\mathbb{R}(\mathbf{M}(n)) & \mathbb{I}(\mathbf{M}(n)) \\
\mathbb{I}(\mathbf{M}(n)) & -\mathbb{R}(\mathbf{M}(n))
\end{bmatrix}
\begin{bmatrix}
\mathbb{R}(\boldsymbol{\phi}_r) \\
\mathbb{I}(\boldsymbol{\phi}_r)
\end{bmatrix} \\
\end{align*}

Therefore, the complex problem $\mathbb{R}(\boldsymbol{\phi}^H\overline{\mathbf{M}}(n)\boldsymbol{\phi})$ has been transformed to a real problem as $\widetilde{\boldsymbol{\phi}}_r^H\widetilde{\mathbf{M}}(n)\widetilde{\boldsymbol{\phi}}_r$. The equation can be further derived by using a second-Taylor expansion as

\begin{equation}
   \begin{aligned}
       &\widetilde{\boldsymbol{\phi}}_r^H\widetilde{\mathbf{M}}(n)\widetilde{\boldsymbol{\phi}}_r\\
       &\leq \widetilde{\boldsymbol{\phi}}_{r,j}^H\widetilde{\mathbf{M}}(n)\widetilde{\boldsymbol{\phi}}_{r,j}+\widetilde{\boldsymbol{\phi}}_{r,j}^H(\widetilde{\mathbf{M}}(n)+\widetilde{\mathbf{M}}^H(n))(\widetilde{\boldsymbol{\phi}}_r-\widetilde{\boldsymbol{\phi}}_{r,j})\\
       &+\frac{\lambda_{\overline{m}_0}(n)}{2}(\widetilde{\boldsymbol{\phi}}_r-\widetilde{\boldsymbol{\phi}}_{r,j})^H(\widetilde{\boldsymbol{\phi}}_r-\widetilde{\boldsymbol{\phi}}_{r,j})\\
       &=\frac{\lambda_{\overline{m}_0}(n)}{2}\boldsymbol{\phi}_r^H\boldsymbol{\phi}_r+\mathbb{R}(\boldsymbol{\phi}_r\overline{\mathbf{m}}(n))-\widetilde{\boldsymbol{\phi}}_{r,j}^H\widetilde{\mathbf{M}}^H(n)\widetilde{\boldsymbol{\phi}}_{r,j}\\
       &+\frac{\lambda_{\overline{m}_0}(n)}{2}\widetilde{\boldsymbol{\phi}}_{r,j}^H\widetilde{\boldsymbol{\phi}}_{r,j},
   \end{aligned}
\end{equation}

\noindent where we have $\overline{\mathbf{m}}=\mathbf{U}(\widetilde{\mathbf{M}}(n)+\widetilde{\mathbf{M}}^H(n)-\lambda_{\overline{m}_0}(n)\mathbf{I}_{2Q})\widetilde{\boldsymbol{\phi}}_{r,j}$, with $\mathbf{U}=[\mathbf{I}_Q, I\mathbf{I}_Q]$, where the imaginary part of a complex number is denoted by the italicized uppercase letter $I$. Also, $\lambda_{\overline{m}_0}(n)$ is the eigenvalue of $\widetilde{\mathbf{M}}(n)+\widetilde{\mathbf{M}}^H(n)$. Then, we can write 
$\mathbf{z}^H\mathbf{M}(n)\mathbf{z}$ as

\begin{equation}
   \begin{aligned}
       &\mathbf{z}^H\mathbf{M}(n)\mathbf{z}\\
       &\leq \frac{\lambda_{\overline{m}_0}(n)}{2}\boldsymbol{\phi}_r^H\boldsymbol{\phi}_r+\mathbb{R}(\boldsymbol{\phi}_r\overline{\mathbf{m}}(n))-\widetilde{\boldsymbol{\phi}}_{r,j}^H\widetilde{\mathbf{M}}^H(n)\widetilde{\boldsymbol{\phi}}_{r,j}\\
   &+\frac{\lambda_{\overline{m}_0}(n)}{2}\widetilde{\boldsymbol{\phi}}_{r,j}^H\widetilde{\boldsymbol{\phi}}_{r,j}+\lambda_{m,0}(n)\mathbf{z}^H\mathbf{z}\\
   &+\mathbf{z}_j^H(\lambda_{m,0}\mathbf{I}_{Q^2}-\mathbf{M}(n))\mathbf{z}_j\\
   &=\frac{\lambda_{\overline{m}_0}(n)}{2}\boldsymbol{\phi}_r^H\boldsymbol{\phi}_r+\mathbb{R}(\boldsymbol{\phi}_r\overline{\mathbf{m}}(n))+const.\label{equ23}
   \end{aligned}
\end{equation}

According to \eqref{eq20} and \eqref{equ23}, the optimization problem can be rewritten as

\begin{equation}
    \min_{\boldsymbol{\phi}_r,\boldsymbol{\phi}_t} \boldsymbol{\phi}_r^H\overline{\mathbf{L}}(n)\boldsymbol{\phi}_r+\mathbb{R}(\boldsymbol{\phi}_r^H\overline{\mathbf{m}}(n)),\label{equ24}
\end{equation}

\noindent where $\overline{\mathbf{L}}(n)=\delta\sigma_v^2\mathbf{L}(n)+\frac{\lambda_{\overline{m}_0}(n)}{2}\mathbf{I}_Q.$ It is clear that \eqref{equ24}
is a convex problem. Hence, we can use tools such as CVX to solve this problem.
\section{Range and velocity estimation}
Optimizing the radar SNR inherently improves the signal quality and enhances the distinction between the target echo and noise, leading to more accurate estimations of range and velocity. A higher radar SNR provides a clearer signal representation, allowing for more precise measurement of phase shifts and Doppler shifts, which are crucial for estimating distance and velocity. Consequently, as the radar SNR is maximized, the system can more effectively filter out inaccuracies and noise, thereby reducing the MSE in range and velocity estimations. In this section, we focus on estimations of both range and velocity.

\subsection{Estimations for range and velocity }
In an OFDM system with an ASTARS, the estimated distance for the $n$th subcarrier $\hat{d}(n)$ can be calculated by considering the phase difference $\Delta \phi$ observed in the received signal across adjacent subcarriers, taking into account the ASTARS-induced phase shifts. This estimated distance $\hat{d}(n)$ represents the total path length, which includes the distance from the DFRC BS to the ASTARS, and from the ASTARS to the target, which can be written as \cite{0069}
\begin{equation}
    \hat{d}(n) = \frac{c \Delta \phi_{\text{eff}}(n)}{2\pi \Delta f },\label{equ25}
\end{equation}
where $\hat{d}(n)$ is the estimated total path length for the $n$th subcarrier, $c$ is the speed of light, $\Delta \phi_{\text{eff}}(n)$ is the effective phase difference incorporating the RIS phase shifts, $\Delta f$ is the subcarrier spacing. Given the ASTARS phase vector for radar sensing $\boldsymbol{\Psi}_r$ which consists of phase shifts $\phi_{q}^{r}$ for each RIS element $q$, the phase difference introduced by the ASTARS between adjacent elements can be computed as
\begin{equation}
    \Delta \phi_{\text{RIS}}(q) = \phi_{q+1}^{r} - \phi_{q}^{r}, \quad \text{for } q = 1, 2, \ldots, Q-1.
\end{equation}

The raw phase difference $\Delta \phi_{\text{raw}}(n)$ at the $n$-th subcarrier can be obtained from the phase of the received signal $\mathbf{y}_r(n)$ as 
\begin{equation}
    \Delta \phi_{\text{raw}}(n)=\arg\left(\mathbf{y}_r(n)\right)-\arg\left(\mathbf{y}_r(n-1)\right),
\end{equation}

\noindent where the raw phase difference includes the phase shifts caused by the ASTARS. Then, the effective phase difference $\Delta \phi_{\text{eff}}(n)$, excluding  the ASTARS-induced phase shifts $\sum_{q=1}^{Q-1} \Delta \phi_{\text{RIS}}(q)$, is given by
\begin{equation}
    \Delta \phi_{\text{eff}}(n) = \Delta \phi_{\text{raw}}(n) - \sum_{q=1}^{Q-1} \Delta \phi_{\text{RIS}}(q).
\end{equation}

This effective phase difference $\Delta \phi_{\text{eff}}(n)$ incorporates the phase shifts introduced by the RIS and is used for subsequent calculations of distance and velocity estimations in ASTARS-assisted OFDM systems. In the ASTARS-assisted OFDM system, the estimated velocity $\hat{v}(n)$ can be derived from the observed effective Doppler shift $f_{d,\text{eff}}$ associated with the target motion and the ASTARS configuration. This velocity estimation considers the relative motion between the DFRC BS, the ASTARS, and the target \cite{0089}, which can be written as
\begin{equation}
    \hat{v}(n) = \frac{f_{d,\text{eff}}(n) \lambda}{2},\label{equ29}
\end{equation}
where $\hat{v}(n)$ is the estimated velocity, $f_{d,\text{eff}}$ is the effective Doppler shift including RIS effects, and $\lambda$ is the wavelength of the carrier frequency. The effective Doppler shift $f_{d,\text{eff}}$ is influenced by the relative motion and the ASTARS modulation of the signal path. To derive the effective Doppler shift $f_{d,\text{eff}}$ from the effective phase difference, we use
\begin{equation}
    f_{d,\text{eff}}(n) = \frac{\Delta \phi_{\text{eff}}(n)}{2\pi T_s},
\end{equation}
where $T_s$ is the duration of an OFDM symbol.

\subsection{MSE for range and velocity estimations}
In the ASTARS-assisted OFDM system, considering the effects of the ASTARS on the signal propagation, the MSE of the distance estimation for $n$th OFDM subcarrier can be expressed as
\begin{equation}
    \text{MSE}_d(n) = \mathbb{E}\left[\left(\hat{d}(n) - d_{\text{true}}(n)\right)^2\right],\label{equ31}
\end{equation}
where  $d_{\text{true}}(n)$ is the true total path length from the DFRC BS, reflected by the ASTARS, to the target.

The MSE of the velocity estimation, incorporating the impact of ASTARS-induced phase shifts on the Doppler shift, can be formulated as
\begin{equation}
    \text{MSE}_v(n) = \mathbb{E}\left[\left(\hat{v}(n) - v_{\text{true}}(n)\right)^2\right],\label{equ32}
\end{equation}
where  $v_{\text{true}}(n)$ is the true velocity of the target relative to the DFRC base station. The whole algorithm for the ASTARS optimization and range/velocity estimations is given as Algorithm $1$.
\begin{figure}[!t]
	\label{alg:LSB}
	\begin{algorithm}[H]
		\caption{ASTARS optimization and range/velocity estimations}
		\begin{algorithmic}[1]
            \STATE \textbf{Input:} Initial ASTARS phase coefficients $\boldsymbol{\Psi}_r^0$ and $\boldsymbol{\Psi}_t^0$, initial transmit beamforming matrix $\mathbf{W}^0$, subcarrier spacing $\Delta f$, wavelength $\lambda$, and OFDM period $T_s$.
            \REPEAT
                \STATE Optimize the transmit beamforming matrix $\mathbf{W}$ according to Eq. \eqref{equ11}.
                \STATE Convert problem $\mathcal{P}3$ to a convex form and optimize ASTARS phase coefficients $\boldsymbol{\Psi}_r$ and $\boldsymbol{\Psi}_t$ using Eq. \eqref{equ24}.
            \UNTIL{The radar SNR is maximized subject to the constraints.}
            \STATE Calculate the range estimate $\hat{d}(n)$ using Eq. \eqref{equ25}.
            \STATE Calculate the velocity estimate $\hat{v}(n)$ using Eq. \eqref{equ29}.
            \STATE Determine the estimation MSEs for range and velocity using Eq. \eqref{equ31} and \eqref{equ32}, respectively.
            \STATE \textbf{Output:} Optimized ASTARS phase coefficients $\boldsymbol{\Psi}_r$ and $\boldsymbol{\Psi}_t$, transmit beamforming matrix $\mathbf{W}$, and MSEs for range $\text{MSE}_d(n)$ and velocity $\text{MSE}_v(n)$.
     \end{algorithmic}
	\end{algorithm}
\end{figure}

\subsection{Complexity of Algorithm $1$}
\textcolor{black}{The complexity of Algorithm $1$ primarily depends on two main optimization steps: the transmit beamforming optimization and the ASTARS phase coefficients optimization. The transmit beamforming optimization involves solving a QSDP for the transmit beamforming matrix 
$\mathbf{W}$, which has a complexity of 
$\mathcal{O}((M^2+MK)^{3.5})$, where the factor $3.5$ reflects the worst-case polynomial time complexity for such problems, taking into account the matrix dimensions of $M$ and $K$. The ASTARS phase coefficients optimization involves converting the optimization problem to a convex form and solving for the phase coefficients $\boldsymbol{\Psi}_r$ and $\boldsymbol{\Psi}_t$, with a complexity of $\mathcal{O}(Q^3)$, where the factor $3$ is derived from the cubic time complexity for solving such problems in terms of the number of ASTARS elements 
$Q$. The total complexity of Algorithm 1 is $\mathcal{O}(((M^2+MK)^{3.5}+Q^3))$. This reflects the computational intensity of the optimization steps involved.}

\section{Simulations}
In this section, we provide simulation results for our proposed ASTARS-aided ISAC systems with OFDM. The DFRC BS is equipped with $M=32$ antennas, which serves $K=4$ users located on the communication area. The ASTARS comprises an equal number of elements in both dimensions, for example, forming a $10 \times 10$ array when $Q=100$ elements. In our system, we assume a carrier frequency $3$ GHz with wavelength $\lambda=0.1$ m. \textcolor{black}{We assume the large-scale fading coefficients
$\widetilde{\beta}_{d,k}=C_1d_1^{-\mu_1}$, $\widetilde{\beta}_{G}=C_2d_2^{-\mu_2}$, where we have $C_1=28$ dB and $C_2=26$ dB. To ensure the path loss coefficients are appropriately scaled and less than 1, we convert these values to a linear scale as $C_1' = 10^{-2.8} \approx 1.58 \times 10^{-3}$ and $C_2' = 10^{-2.6} \approx 2.51 \times 10^{-3}$, maintaining the same path loss characteristics.} Moreover, $\mu_1=3.67$ and $\mu_2=2.2$ are path-loss components. $d_1$ and $d_2$ are the distances of BS-users and BS-RIS with $d_1=50$ m and $d_2=8$ m. Also, we assume that $\eta_k$ and $\widetilde{\beta}_a$ have the same value as $\widetilde{\beta}_{d,k}$. Excluding the comparison of varying values for these two parameters, it is specified that the system employs $N_s=64$ subcarriers with a subcarrier spacing of $\Delta f=120$ kHz (5G NR standard). The duration of an OFDM symbol is $T_s=17.68\,\mu $s. The noise powers are given by $\sigma^2=\sigma_v^2=-40$ dB and $\sigma_r^2=-58$ dB. The maximum value of AAC is $\alpha_{max}=4$. The dimensions for each element within ASTARS are set as $d_H=d_V=\lambda/4$. The channel $\mathbf{G}$ is based on Rician fading with Rician factor $R=10$ dB. We apply the method described in Sec. III to optimize the transmit beamforming and phase coefficients of the ASTARS. By using the results from this optimization, we depict the radar SNR and perform estimation of range and velocity. We assume that the unit of range is m (meters) and the unit of velocity is km/h (kilometers per hour).

\begin{figure}[t]
\flushleft \includegraphics[width=0.5\textwidth]{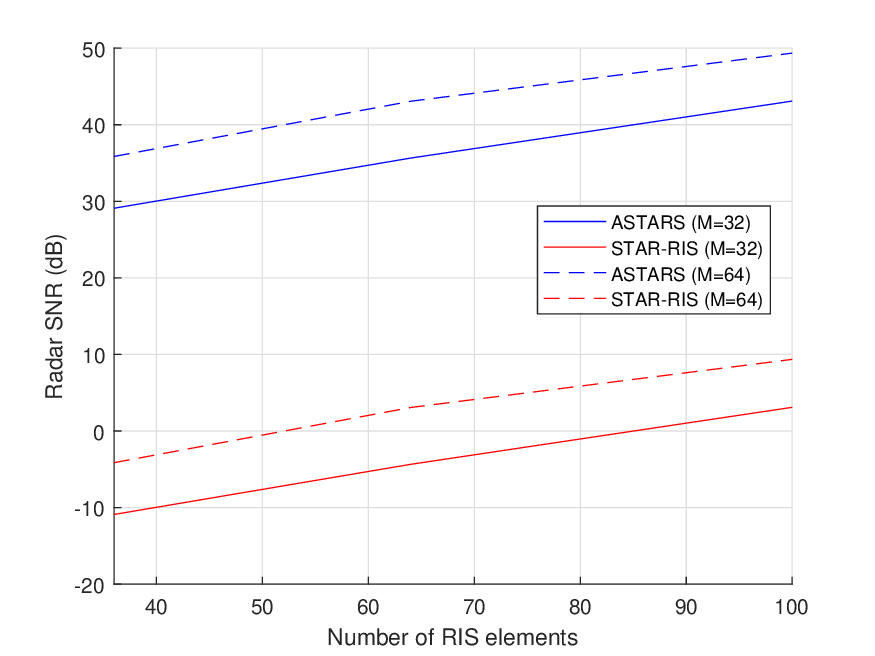} \caption{Radar SNR versus the number of ASTARS elements for ASTARS and STAR-RIS.}
\end{figure}

\begin{figure}[t]
\flushleft \includegraphics[width=0.5\textwidth]{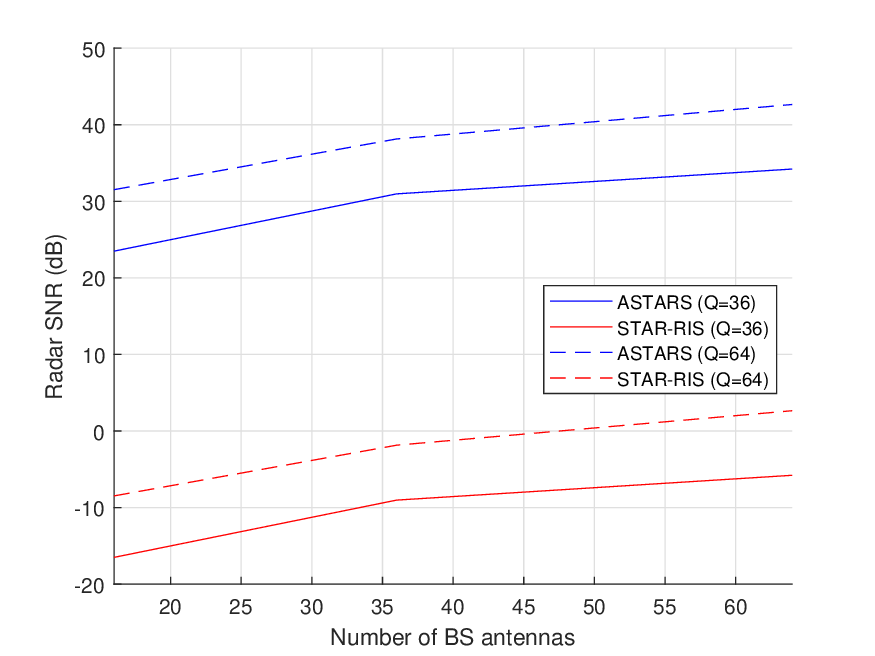} \caption{Radar SNR versus the  number of DFRC BS antennas for ASTARS and STAR-RIS.}
\end{figure}

\begin{figure}[t]
\flushleft \includegraphics[width=0.5\textwidth]{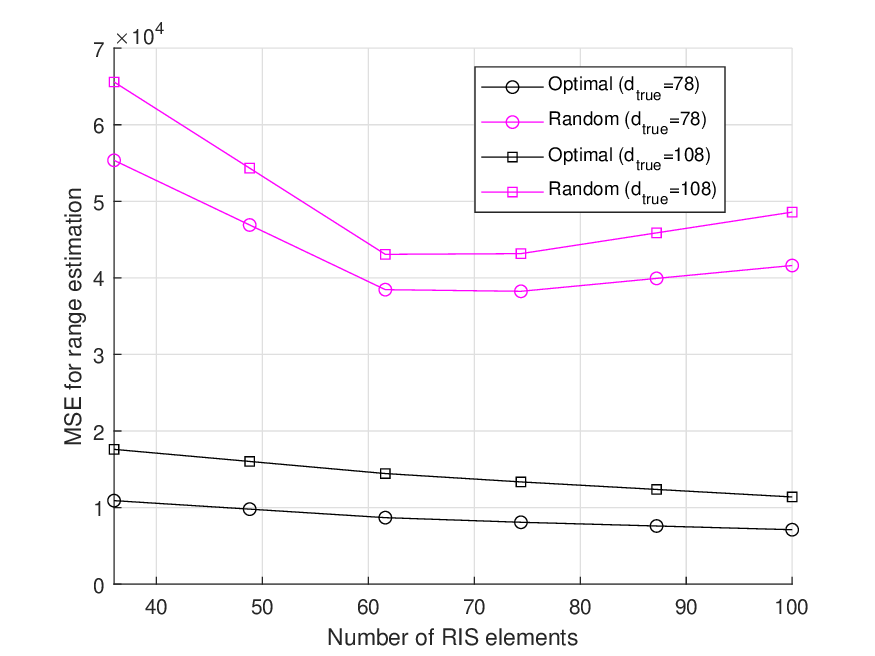} \caption{Comparison of optimal phases/amplitudes with the random case for range estimation.}
\end{figure}

\begin{figure}[t]
\flushleft \includegraphics[width=0.5\textwidth]{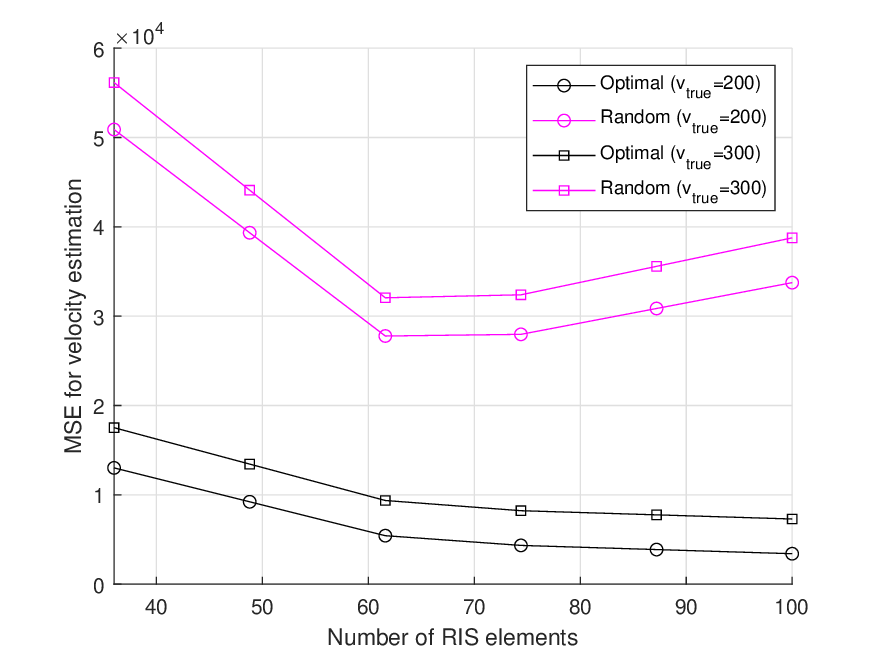} \caption{Comparison of optimal phases/amplitudes with the random case for velocity estimation.}
\end{figure}

\begin{figure}[t]
\flushleft \includegraphics[width=0.5\textwidth]{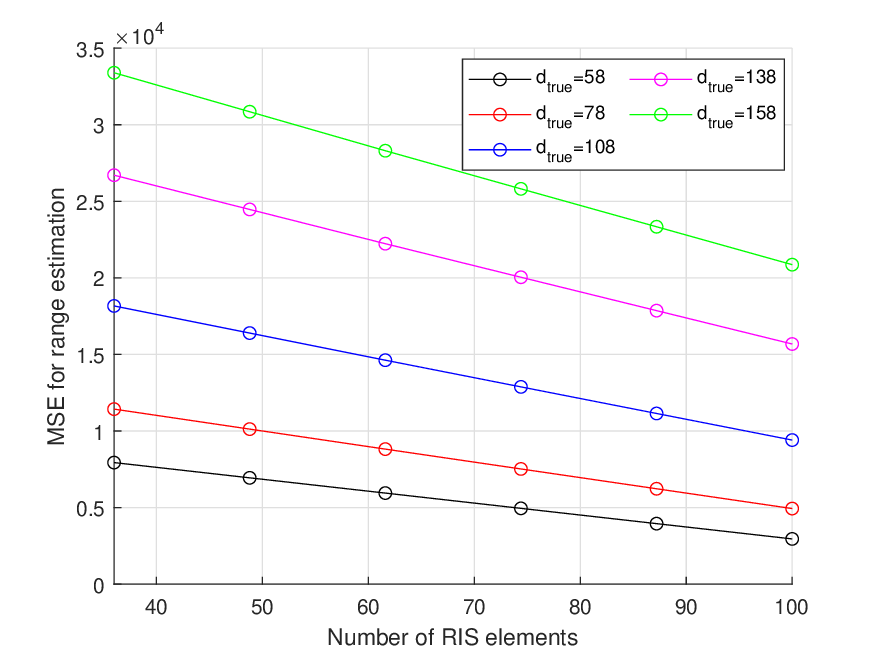} \caption{Range estimation MSE versus the number of RIS elements.}
\end{figure}

\begin{figure}[t]
\flushleft \includegraphics[width=0.5\textwidth]{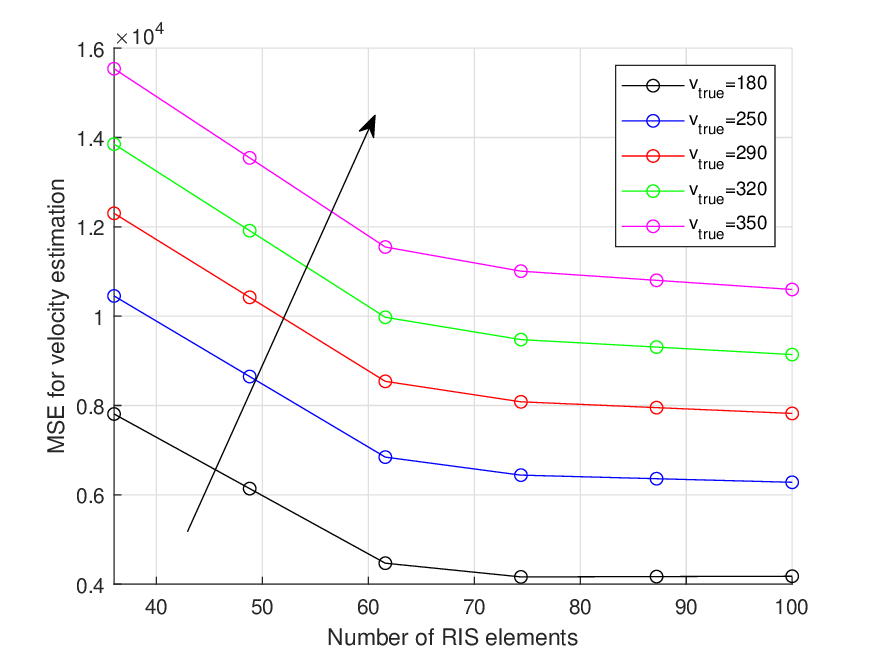} \caption{Velocity estimation MSE versus the number of RIS elements for different $v_{\textup{true}}$.}
\end{figure}

\begin{figure}[t]
\flushleft \includegraphics[width=0.5\textwidth]{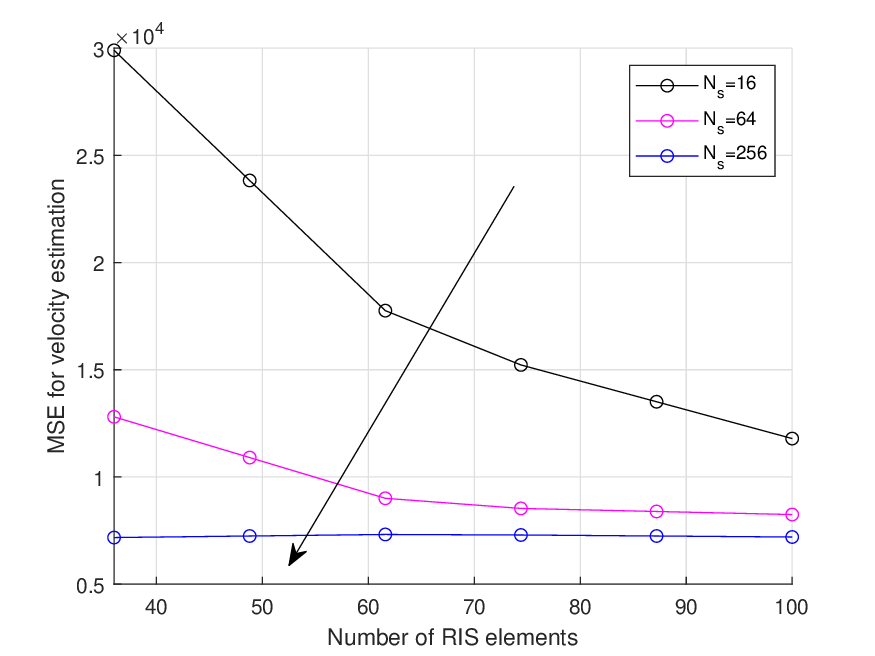} \caption{Velocity estimation MSE versus the number of RIS elements for different numbers of subcarriers $N_{s}$.}
\end{figure}

\begin{figure}[t]
\flushleft \includegraphics[width=0.5\textwidth]{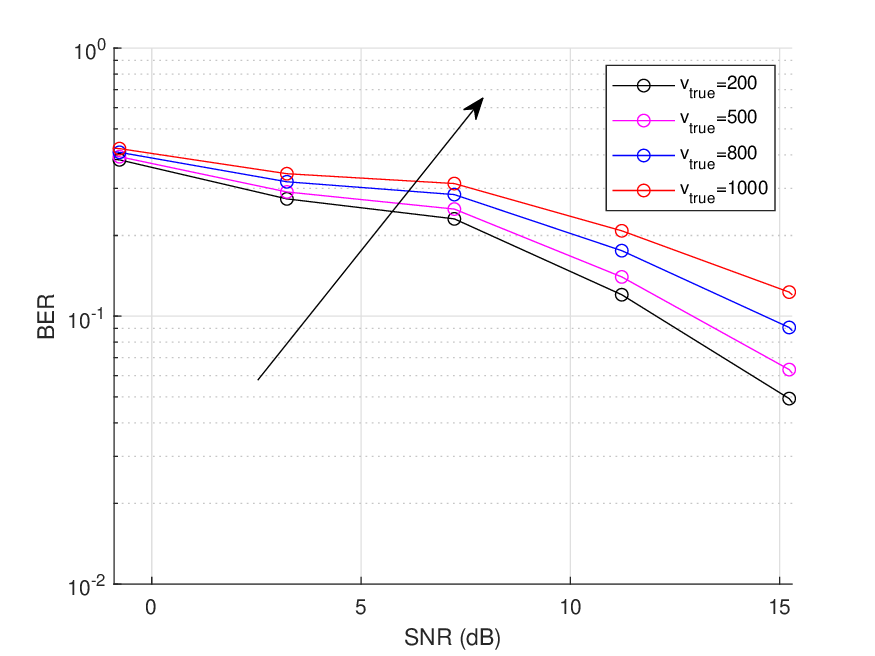} \caption{\textcolor{black}{BPSK BER versus the transmit SNR for different  $v_{\textup{true}}$  ($\Delta f=60$ kHz)}.}
\end{figure}

\begin{figure}[t]
\flushleft \includegraphics[width=0.5\textwidth]{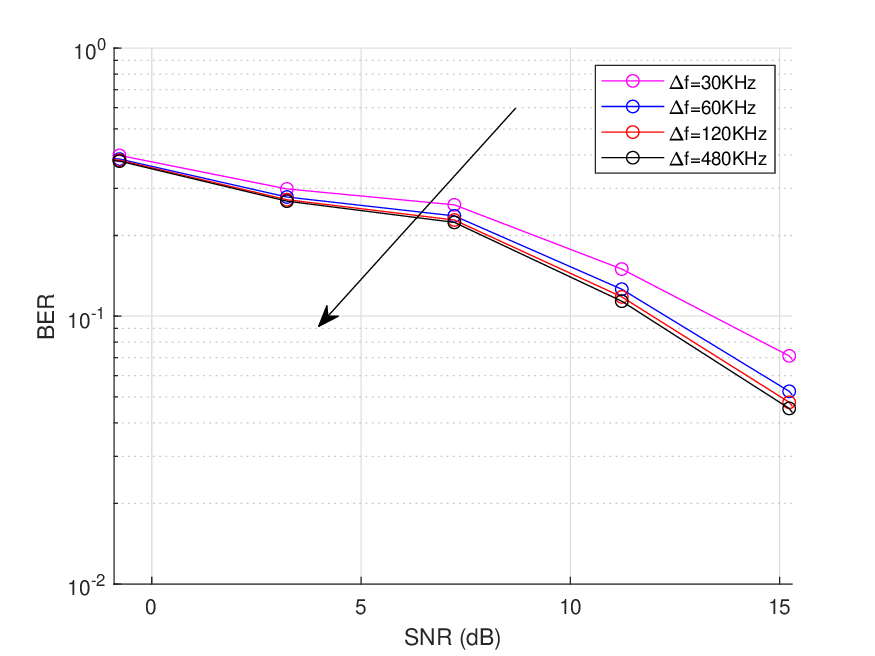} \caption{BPSK BER versus the transmit SNR for different  subcarrier spacing $\Delta f$ ($v_{\textup{true}}=300$ km/h).}
\end{figure}

\begin{figure}[t]
\flushleft
\includegraphics[width=0.5\textwidth]{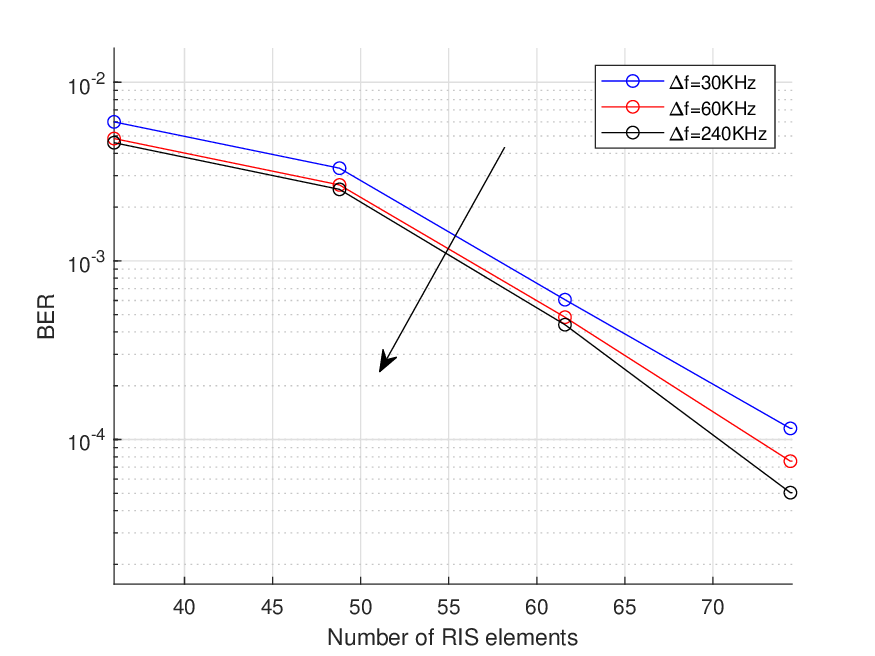} \caption{BPSK BER versus the number of RIS elements for different subcarrier spacing $\Delta f$ ($v_{\textup{true}}=300$ km/h, SNR$=20$ dB).}
\end{figure}

\subsection{Radar SNR versus number of RIS elements or DFRC BS antennas}
In Figs. $2$ and $3$, we have plotted the radar SNR against the number of ASTARS elements and DFRC BS antennas by solving the QSDP problem outlined in Sec. III. It is clear that when we increase the number of ASTARS panels, the radar SNR increases. This improvement can be attributed to the enhanced reflective capabilities of a larger ASTARS panel, which effectively amplifies the signal reflected back to the radar receiver. With more panels, the ASTARS can reflect more beams towards the target, thereby increasing the signal power at the receiver. Additionally, a larger ASTARS array can better mitigate the interference in our system, further boosting the received signal strength and enhancing the radar SNR. For the number of DFRC BS antennas, we can also find that the radar SNR increases when the number of BS antennas increases. This is due to the beamforming gain achieved with a larger antenna array at the DFRC BS. As the number of antennas increases, the BS can transmit more beams to the ASTARS, which results in a more concentrated energy transmission. Additionally, a larger antenna array provides better spatial resolution and diversity, which further enhances the capability of the system to distinguish between different targets and interference, leading to improved radar detection.

\subsection{Radar SNR for ASTARS and STAR-RIS}
In Figs. $2$ and $3$, we have also compared the performance of ASTARS and STAR-RIS in the ISAC system. If we ignore the ACC and the dynamic noise, we can get the structure of the STAR-RIS. The simulation results show that the ASTARS can improve the radar SNR. This is because, within this framework, ASTARS leverages a minimal amount of power to enhance the strength of the received signal. Unlike the traditional STAR-RIS, which only reflects the incoming signal, ASTARS incorporates active elements that can add additional energy into the reflected signal. This active amplification enables ASTARS to significantly boost the strength of the signal received at the radar.

\subsection{Range and velocity MSE for random and optimal ASTARS}

Although our optimization approach in Sec. III primarily aims to maximize the radar SNR, it also enhances range and velocity estimation performance in the sensing domain. It is evident that our optimized ASTARS configuration results in reduced MSE for both range and velocity estimations in Figs. $4$ and $5$. Moreover, as we increase the number of ASTARS elements, the MSEs for both estimations consistently decrease for our optimized case. Conversely, in the random case (\textcolor{black}{RIS phase is random from $-\frac{\pi}{2}$ to $\frac{\pi}{2}$}), MSEs initially drop but subsequently rise, indicating the optimized system capacity to stabilize estimations. This stability is because the optimized ASTARS can concentrate the signal energy more effectively towards the target, thereby improving the measurement accuracy of the signals. By optimizing for radar SNR, we inherently enhance the system ability to distinguish between the signal and noise, which is important for accurate range and velocity estimations. Essentially, a higher radar SNR implies a signal reflection from the target, which directly contributes to more precise sensing outcomes.

\subsection{Range and velocity estimations performance analysis}
The MSE for range and velocity estimations versus number of ASTARS elements has been shown in Figs. $6$ and $7$.  We can find that when we increase the elements number, the MSE decreases at the same time. This is because a more large ASTARS array allows for a more precise manipulation of the phase and amplitude of the reflected signals, thereby improving the accuracy of the phase measurements used in range and velocity calculations. Hence, the estimation performance can also been improved. In Figs. $6$ and $7$, we can also find that the MSEs for range and velocity estimations increase when their true values increase. Larger distances or higher velocities introduce greater phase shifts across the signal path. Also, as the target moves farther away or at a faster velocity, the signal undergoes more significant changes which can amplify the estimation errors and system imperfections. 

\subsection{The effect of different number of OFDM subcarriers in the velocity estimation}
In Fig. $8$, we observe a reduction in the velocity estimation MSE as the number of OFDM subcarriers increases. This trend suggests that a higher number of subcarriers helps to reduce the impact of frequency-selective fading, thereby increasing the robustness and accuracy of velocity estimation. Furthermore, while an increased subcarrier density significantly enhances estimation performance, it is important to note that beyond a certain threshold, the rate of improvement in MSE begins to taper off due to practical constraints related to processing power and computational complexity.

\subsection{BPSK BER and high-velocity circumstances}
In Fig. $9$, we compare the binary phase-shift keying (BPSK) BER due to inter-carrier interference (ICI) across different true velocities. The generation of ICI in our simulation is achieved by introducing phase and amplitude errors that scale with the Doppler frequency shift. Specifically, the phase error is modeled as a random variable uniformly distributed between 
$-\pi$ and $\pi$, scaled by an ICI effect factor. The amplitude error is modeled as a small perturbation scaled by the square root of the ICI effect factor. These errors are applied to each subcarrier, simulating the ICI experienced in real-world high-speed scenarios. It is clear that as $v_{\textup{true}}$ increases, the BER also increases. This indicates that the data estimation performance deteriorates under high-velocity conditions. Consequently, in Fig. $10$, we adhere to the 5G NR criteria, observing that an increase in subcarrier spacing results in a decrease in BER. This suggests that the data estimation performance in high-speed conditions can be improved by increasing the subcarrier spacing. 
Increasing the subcarrier spacing reduces the impact of Doppler shifts, which are more pronounced at higher velocities. This mitigation of Doppler effects is crucial for maintaining the integrity of signal transmission, also enhancing data estimation performance in high-speed scenarios. In Fig. $11$, we have plotted the number of RIS elements versus BPSK BER. The results show that when we increase the number of RIS elements, the BPSK BER decreases simultaneously. This occurs because more RIS elements enhance the signal reflection and focusing capabilities, leading to a stronger and more coherent signal at the receiver. Consequently, the improved signal quality reduces the BER, resulting in more reliable communication. Additionally, we can confirm that when the subcarrier spacing increases, the BER decreases at the same time, even when the horizontal axis represents the number of RIS elements. This is because larger subcarrier spacing reduces inter-carrier interference, further improving signal clarity. These combined effects demonstrate the importance of optimizing both RIS configuration and subcarrier spacing for minimizing BER.


\section{Conclusion}
In this work, we have proposed an ASTARS-aided ISAC system. We optimized the transmit beamforming and ASTARS phase coefficients to maximize the radar SNR. We employed OFDM in our system and addressed frequency-selective fading problem by using multiple OFDM subcarriers. We used our optimized ASTARS to carry out range and velocity estimations. The results have shown that when more ASTARS elements are equipped, the estimation MSEs decrease. We have also investigated that we can increase the subcarrier spacing in a high-speed condition to reduce the transmit BER. In our future work, we plan to use orthogonal time frequency space (OTFS) and then analyze the performance for our system \cite{0082,0083,0084}. We will also combine our proposed ISAC system with stacked RIS \cite{0092}.

\appendix

\subsection{Constraint on communication SINR }
In problem $\mathcal{P}1$, we know that the optimization needs to meet the minimum SINR for each communication user. Hence from \eqref{eq6} and \eqref{9a}, we can write

\begin{equation}
    \begin{aligned}
       &|\mathbf{u}_k^H(n)\mathbf{w}_{c,k}|_2^2\\
       &\geq \xi \big(|\mathbf{u}_k^H(n)\mathbf{W}_r|_2^2+\sum_{i\neq k}\mathbf{u}_k^H(n)\mathbf{R}_i\mathbf{u}_k(n)\\
       &+\sigma_v^2\mathbf{t}_k^H(n)\boldsymbol{\Psi}_{t}\boldsymbol{\Psi}_{t}^H\mathbf{t}_k(n)+\sigma^2\big)\\
       &=\xi A(n).\label{equ33}
    \end{aligned}
\end{equation}

From \eqref{equ33}, we use the equation $A(n)=|\mathbf{u}_k^H(n)\mathbf{W}_r|_2^2+\sum_{i\neq k}\mathbf{u}_k^H(n)\mathbf{R}_i\mathbf{u}_k(n)+\sigma_v^2\mathbf{t}_k^H(n)\boldsymbol{\Psi}_{t}\boldsymbol{\Psi}_{t}^H\mathbf{t}_k(n)+\sigma^2$ to simpify the term on the right side
\begin{equation}
    \frac{A(n)}{\mathbf{u}_k^H(n)\mathbf{R}_k\mathbf{u}_k(n)}\leq \frac{1}{\xi}.
\end{equation}

If we add plus one on each side, we can get

\begin{equation}
    \frac{\mathbf{u}_k^H(n)\mathbf{R}\mathbf{u}_k(n)+\sigma_v^2\mathbf{t}_k^H(n)\boldsymbol{\Psi}_{t}\boldsymbol{\Psi}_{t}^H\mathbf{t}_k(n)+\sigma^2}{\mathbf{u}_k^H(n)\mathbf{R}_k\mathbf{u}_k(n)}\leq \frac{1}{\xi}+1.
\end{equation}

 Then, we can get the constraint \eqref{9a} for maximizing the radar SNR as
\begin{equation}
\begin{aligned}
    &\quad(1+\xi^{-1})\textup{tr}(\mathbf{u}_k(n)\mathbf{u}_k^H(n)\mathbf{R}_k)\\
     &\geq\sigma^2+\textup{tr}(\mathbf{u}_k(n)\mathbf{u}_k^H(n)\mathbf{R}+\sigma_v^2\mathbf{t}_k^H(n)\boldsymbol{\Psi}_{t}\boldsymbol{\Psi}_{t}^H\mathbf{t}_k(n)),\forall k.
\end{aligned}
\end{equation}

\bibliographystyle{IEEEtran.bst}
\bibliography{ST_IEEE2.bib}
\vspace{12pt}

\end{document}